\documentclass[preprint,11pt,authoryear]{elsarticle}

\usepackage{lmodern,sansmath}
\usepackage{textcomp}
\usepackage{microtype}
\usepackage{amsmath}
\usepackage{amssymb}
\usepackage{booktabs}
\usepackage{graphicx}
\usepackage{xcolor}
\usepackage{tikz}

\def\dd#1#2{\frac{\mbox{d} #1}{\mbox{d} #2}}
\def\pp#1#2{\frac{\partial #1}{\partial #2}}
\renewcommand{\vec}[1]{\boldsymbol{#1}}

\newcommand{\avg}[1]{\langle{#1}\rangle}

\journal{IJHFF}

\begin{document}

\begin{frontmatter}



  \title{\textbf{One-point and two-point statistics of\\
      homogeneous isotropic decaying turbulence\\ with variable
      viscosity}}


\author{M.\ Gauding, L.\ Danaila and E.\ Varea}

\address{CORIA -- CNRS UMR 6614, 76801 Saint Etienne du Rouvray,
  France}

\begin{abstract}
  The decay of homogeneous isotropic turbulence in a variable
  viscosity fluid with a viscosity ratio up to 15 is analyzed by means
  of highly resolved direct numerical simulations (DNS) at low
  Reynolds numbers. The question addressed by the present work is how
  quantities such as the kinetic energy and the associated dissipation
  rate, as well as the inter-scale transport mechanism of turbulence
  are changed by local fluctuations of the viscosity. From the
  one-point budget equation of the turbulent kinetic energy, it is
  shown that the mean dissipation rate is nearly unchanged by variable
  viscosity effects. This result is explained by a negative
  correlation between the local viscosity and the local velocity
  gradients. However, the dissipation is a highly fluctuating quantity
  with a strong level of intermittency. From a statistical analysis it
  is shown that turbulent flows with variable viscosity are
  characterized by an enhanced level of small-scale intermittency,
  which results in the presence of smaller length scales and a
  modified turbulent mixing.  The effect of variable viscosity on the
  turbulent cascade is analyzed by a budget equation for the velocity
  structure function.  From DNS it is shown that viscosity
    gradients contribute to the inter-scale transport mechanism in the
    form of an inverse transport, where information propagates from
    the small scales to the large scales.
\end{abstract}


\end{frontmatter}

\section{Introduction}
Turbulent flows encountered in engineering and environmental
applications are very often characterized by spatio-temporal
fluctuations of viscosity, which results from variations of
temperature or species composition. A prominent example from
geophysical flows is the convection in the earth's mantle, where the
viscosity decreases with temperature. An other important case is the
turbulent mixing in combustion systems, where a concentration
dependent viscosity may affect the efficiency of turbulent mixing.

Fully developed turbulence reveals a large range of length scales,
varying from the so-called integral length scale $l_t$, at which large
velocity fluctuations occur on average, down to the smallest scale,
the so-called Kolmogorov or dissipation scale $\eta$, at which
turbulent fluctuations are dissipated due to viscosity. According to
Kolmogorov's first hypothesis
(\cite{kolmogorov1941dissipation,kolmogorov1941local}), the statistics
of the smallest scales should be universal, and depend only on two
parameters, namely the viscosity $\nu$ and the mean energy dissipation
rate $\avg{\varepsilon}$.  Kolmogorov's second hypothesis postulates
that large scales of the flow decouple from the smallest scales and
should become independent of viscosity, provided that the Reynolds
number is sufficiently high.  However, numerous experimental and
numerical studies have indicated that Kolmogorov's traditional view is
a crude assumption and that large and small scale quantities are
strongly coupled, cf.~\cite{sreenivasan1997}, \cite{warhaft2000}, and
\cite{antonia2015boundedness}.  The situation is even more complex for
turbulent mixing with local viscosity variations. One has to cope with
a turbulence-scalar interaction which is two-fold: the fluid motions
affect the scalar mixing, while mixing induced viscosity changes
affect the dynamics of the velocity field.

Viscosity represents the most important property of turbulent
  flows, and the impact of its variation on the dynamics should be
  addressed in detail. Most studies reported in literature have
  focused on the impact of variable viscosity on the large
  scales. \citet{chhabra2005entrainment} and
  \citet{talbot2013variable} studied turbulent jet flows, where the
  viscosity between jet fluid and host fluid differs. They observed,
  compared to a flow with uniform viscosity, that the relationship
  between production and dissipation of turbulent energy is altered
  and that the entrainment and the spreading rates of the jet flow are
  changed. An altered spreading rate was also observed by direct
  numerical simulations of turbulent shear layers with variable
  viscosity by \cite{taguelmimt2016effect,taguelmimt2016effects}
  indicating the ability of viscosity variations to modify the largest
  scales of the flow.  \cite{voivenel2017variable} derived generalized
  scale-by-scale budget equations for velocity increments in
  inhomogeneous and anisotropic turbulence with variable
  viscosity. Based on this work, \citet{danaila2017self} showed that
  variable viscosity effects can invalidate the self-similarity of
  turbulent jet flows.  The analysis of \citet{lee2008validity}
  focused on the turbulent mixing of two initially segregated fluids
  with different viscosity in homogeneous isotropic turbulence. They
  found that the dissipation becomes independent of viscosity
  confirming Taylor's postulate.  \cite{taylor1935statistical}
  postulated that the mean energy dissipation $\avg{\varepsilon}$
  depends only on the large-scale velocity fluctuations $u'$ and the
  integral length scale $l_t$, i.e.\
  \begin{equation}
    \label{eq:tay1}
  \avg{\varepsilon} \propto \frac{u'^3}{l_t} \,,
\end{equation}
and hence becomes independent of viscosity, provided that the Reynolds
number is sufficiently large. In eq.~\eqref{eq:tay1}, the
characteristic large-scale velocity fluctuations are defined as
\begin{equation}
  u'=\sqrt{\avg{u_i^2}/3} \,,
\end{equation}
and the integral length scale $l_t$ is defined as
\begin{equation}
  l_t = \frac{3\pi}{4} \frac{
    \int \kappa^{-1} E(\kappa,t) {\rm d} \kappa
  }{
    \int E(\kappa,t) {\rm d} \kappa
  }\,,
\end{equation}
with $E(\kappa)$ being the three-dimensional energy spectrum and
$\kappa$ the magnitude of the wave-number vector.  Ensemble-averages
are denoted by angular brackets and Einstein's summation convention is
used, which implies summation over indices appearing twice.

In this paper, we investigate the decay of homogeneous isotropic
turbulence in a variable viscosity fluid by means of highly resolved
direct numerical simulations (DNS). We address the question how
quantities such as the kinetic energy and the associated dissipation
rate, as well as the inter-scale transport mechanism of turbulence are
changed by local fluctuations of the viscosity.  The paper is
structured as follows. Section 2 presents the governing equations and
the direct numerical simulations on which the analysis is
based. Section 3 introduces the budget equation of the turbulent
energy and discusses the impact of variable viscosity on the
dissipation mechanism of turbulence. Section 4 addresses the impact of
variable viscosity on the viscous cut-off scales of turbulence. An
analysis of the inter-scale transport mechanism based on a budget
equation for the second-order velocity structure function is presented
in section 5. We summarize this study in section 6.

\section{Direct numerical simulations and governing equations} 
  Direct numerical simulations of homogeneous isotropic turbulence
  with variable viscosity at three different viscosity ratios between
  1 and 15 have been performed.  The DNS solves the three-dimensional
  incompressible Navier-Stokes equations,
\begin{equation}
  \label{eq:ns}
  \pp{u_j}{t} + u_i \pp{u_j}{x_i} = -\pp{p}{x_j} +
  \pp{}{x_i}\left(2\nu s_{ij} \right) \,,
\end{equation}
with the continuity equation
\begin{equation}
  \label{eq:konti}
  \pp{u_i}{x_i} = 0 \,,
\end{equation}
in a triply periodic box with size~$2\pi$ by a pseudo-spectral method.
In eqs. \eqref{eq:ns} and \eqref{eq:konti}, the velocity field is
denoted by $u_j$, $p$ is the pressure (for simplicity the density
$1/\rho$ is incorporated in $p$), $\nu$ is the local viscosity, and
$s_{ij}$ is the strain-rate tensor, defined as
\begin{equation}
  s_{ij} = \frac{1}{2} \left( \pp{u_j}{x_i} + \pp{u_i}{x_j} \right) \,.
\end{equation}
  
The local viscosity field $\nu(\vec x, t)$ is determined by solving an
advection-diffusion equation for a scalar field $\phi(\vec x,t)$,
i.e.\
\begin{equation}
  \label{eq:ps}
  \pp{\phi}{t} + u_i \pp{\phi}{x_i} = D \pp{^2\phi}{x_i^2} \,.
\end{equation}
The scalar $\phi$ is statistically isotropic and homogeneous. For
numerical convenience the scalar is further bounded, i.e.\
$-1\le \phi(\vec x,t) \le 1$, and has zero mean, i.e.\
$\avg{\phi}=0$. Ensemble-averages are computed due to homogeneity over
the full computational domain. The local viscosity $\nu(\vec x,t)$ is
linked through a linear relation to the scalar field,
e.g. \cite{grea2014effects}
\begin{equation}
  \label{eq:nu}
  \nu(\vec x,t) = \avg{\nu}+ \nu'(\vec x,t) = \avg{ \nu } + c
  \phi(\vec x,t) \,,
\end{equation}
where $\avg \nu$ denotes the uniform mean viscosity, $\nu'(\vec x,t)$
denotes the fluctuating viscosity field, and $c$ is a positive
constant with $c<\avg \nu$ to ensure positivity of $\nu(\vec x,t)$. A
linear relation between $\nu$ and $\phi$ is convenient because it
keeps the mean viscosity $\avg \nu$ unchanged during the decay.  The
constant $c$ is obtained from the initial minimum and maximum values
of the viscosity by $c=(\nu_{\rm max} - \nu_{\rm min})/2$, which
implies that the initial scalar variance $\avg{\phi^2}$ equals unity.
In the following, we use $\phi$ as a surrogate for $\nu$.  The
molecular diffusivity $D$ in eq.\ \eqref{eq:ps} is assumed to be
constant and equals the mean viscosity
$\avg\nu=(\nu_{\rm min} + \nu_{\rm max})/2$. As a consequence, the
Schmidt number, defined as $\mathit{Sc}=\nu/D$, is a fluctuating
quantity.

The following paragraph briefly summarizes the main features of the
DNS. More details about the numerical procedure and the
parallelization approach are given by
\cite{gauding2015line,gauding2017high}.   Adapting the
approach by \cite{mansour1994decay}, the Navier-Stokes equations are
formulated in spectral space as
\begin{equation}
  \label{eq:nss}
  \pp{}{t} \left( \hat u_i \exp(\nu \kappa^2 t) \right) = \exp(\nu
  \kappa^2 t) P_{ij} \hat H_j \,,
\end{equation}
where 
\begin{equation}
\hat H_j   = - i \kappa_i \mathcal F \left(  u_i u_j -   2 c
    \phi(\vec x,t) s_{ij} \right) 
\end{equation} 
is the Fourier transform of the non-linear terms, including the
convective term and the non-linear part of the viscous term. The
wave-number vector is denoted by $\vec \kappa$, and the Fourier
transform of the velocity field is denoted by $\hat {\vec u}$.  The
projection operator
$P_{ij} = \delta_{ij} - \kappa_i \kappa_j/\kappa^2$ imposes
incompressibility.  The non-linear terms are computed in
physical space and a truncation technique with a smooth spectral
filter is applied to remove aliasing errors. The smooth spectral
filter is highly localized in both real and spectral space, and was
found to be dynamically very stable, cf.~\cite{hou2007computing}. This
feature is relevant for the present study to prevent instabilities at
high wave-number modes caused by the non-linear part of the viscous
term.  An integrating factor technique is used for an exact
integration of the linear part of the viscous terms. Temporal
integration is performed by a low-storage, stability preserving,
third-order Runge-Kutta scheme.  An additional necessary constraint
that has to be satisfied by the DNS is an adequate resolution of the
smallest scales. As proposed by \citet{mansour1994decay}, we require
that for all times, the condition $\kappa_{\rm max} \eta_0 \ge 1$ is
satisfied, where $\eta_0=(\avg{\nu}^3/\avg{\varepsilon})^{1/4}$ is the
Kolmogorov length scale and $\kappa_{\rm max}$ is the largest resolved
wave-number. A grid resolution of $1024^3$ points is used to
appropriately account for both small and large scales.

 Let us now turn our attention to the initialization of the DNS.
  For the velocity field, we follow the approach of
  \cite{ishida2006decay} and prescribe a broad-band energy spectrum
of the type
\begin{equation}
  \label{eq:spec}
E(\kappa,t=0) \propto \kappa^4 \exp (-2 (\kappa/\kappa_p)^2) \,.
\end{equation}
The peak of the initial spectrum is located at $\kappa_p=12$ to inject
energy at sufficiently large length scale while keeping the initial
integral length scale~$l_t$ small compared to the confinement imposed
by the computational domain. The initialization guarantees that
turbulence is statistically homogeneous, isotropic, and
incompressible.  The initial Reynolds number, defined with the mean
viscosity as $\mathit{Re}_0 = u'/(\kappa_p \avg\nu)$, equals 43.

In this work, we compare three different DNS.  The baseline case has a
constant and spatially uniform viscosity of
$\avg\nu=0.006$. Additionally, two cases with variable viscosity are
considered. These cases have the same mean viscosity $\avg\nu$ as the
baseline case, but an initial viscosity ratio
$R_\nu=\nu_{\rm max}/\nu_{\rm min}$ that equals 5 and 15,
respectively. The initial viscosity distribution is bi-modal with an
initial normalized viscosity variance of
$\avg{\nu'^2}/\avg{ \nu}^2 = (\nu_{\rm max} - \nu_{\rm
  min})^2/(\nu_{\rm max} + \nu_{\rm min})^2 $.  At later times, the
viscosity distribution is smoothed due to turbulent mixing, so that
the viscosity variance $\avg{\nu'^2}$ decreases. The initial scalar
energy spectrum is proportional to the velocity spectrum, but the
initial scalar field is not correlated with the velocity field, i.e.\
$\avg{u_i\phi}=0$.  As the mean viscosity $\avg{\nu}$ stays constant
during the decay, the differences between the cases can be attributed
solely to the viscosity ratio $R_\nu$.

\section{Budget of the one-point turbulent kinetic energy}
\label{sec:2}
\begin{figure}
  \centering
  \includegraphics[width=0.65\linewidth]{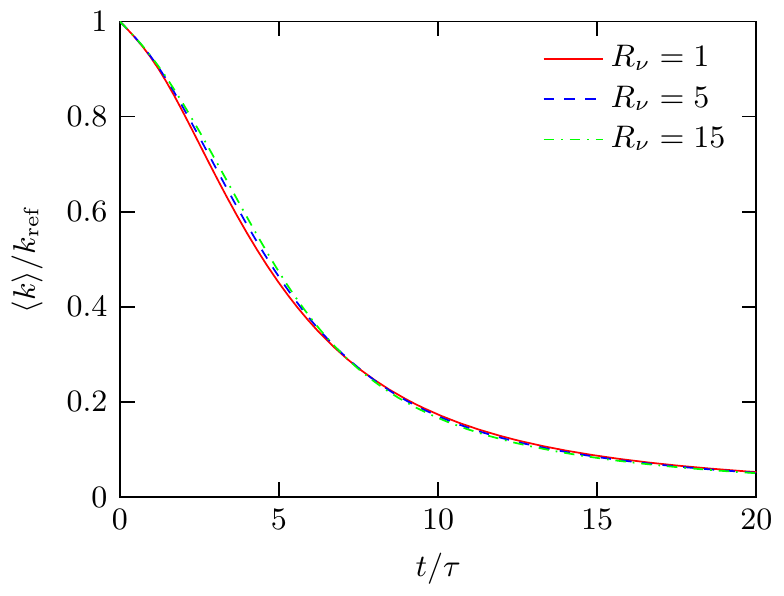} \\
  \includegraphics[width=0.65\linewidth]{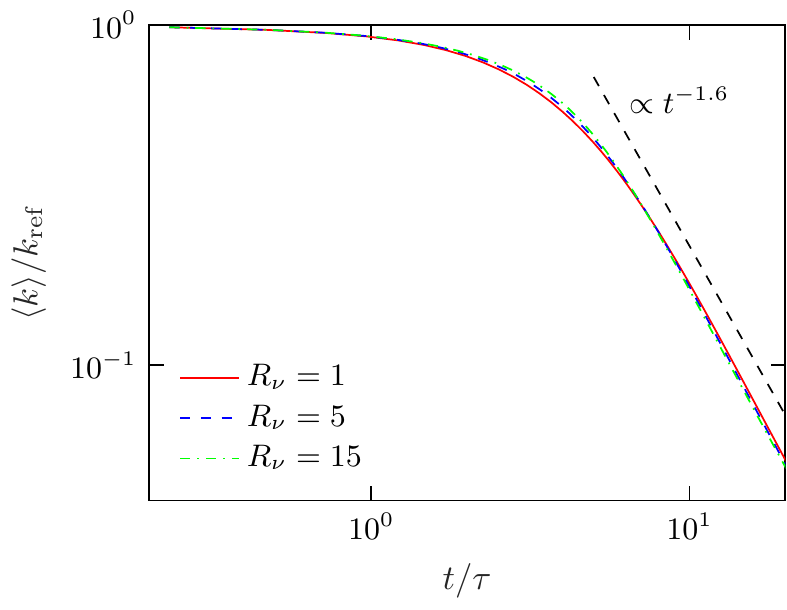} \\ 
  \caption{Temporal evolution of the mean turbulent kinetic energy
    $\avg{k}$ for the three viscosity ratios $R_\nu$. The time is
    normalized by $\tau=1/(\kappa_p u'_{0})$, and
    $k_{\rm ref}=\avg{k}(t=0)$.  }
  \label{fig:te}
\end{figure}

An important question addressed by the present work is how the
dissipation mechanism of turbulence is changed by variable viscosity,
and whether turbulent flows with variable viscosity dissipate more or
less energy as flows with constant viscosity. To answer this question,
we first examine the transport equation for the mean turbulent kinetic
energy $\avg{k}=\avg{u_iu_i}/2$, which reads in decaying homogeneous
isotropic turbulence
\begin{equation}
  \label{eq:ed0}
  \dd{\avg{k}}{t} = \left\langle \pp{\nu}{x_i} \pp{}{x_j} \left(u_iu_j\right) \right\rangle -\avg{\varepsilon}_{\rm VV} \,.
\end{equation}
The first term on the right-hand side describes the dissipation of
turbulent energy due to viscosity gradients. This term is negligible
compared to the second term, which is the mean dissipation rate
\begin{equation}
\left\langle\varepsilon\right\rangle_{\rm VV}=\left\langle\nu \left( \pp{u_i}{x_j}\right)^2 \right\rangle
\end{equation}
of the turbulent kinetic energy $\avg{k}$. Figure~\ref{fig:te} shows
the temporal evolution of the mean turbulent kinetic energy
$\avg{k}$. The mean turbulent kinetic energy is virtually unaffected
by variable viscosity effects, and only a slightly reduced initial
decay is visible for the cases with $R_\nu>1$. These results confirm
the trends previously reported by
\cite{taguelmimt2016effect,taguelmimt2016effects} in DNS studies of
temporally evolving mixing layers.

\begin{figure}
  \centering
  \includegraphics[width=0.65\linewidth]{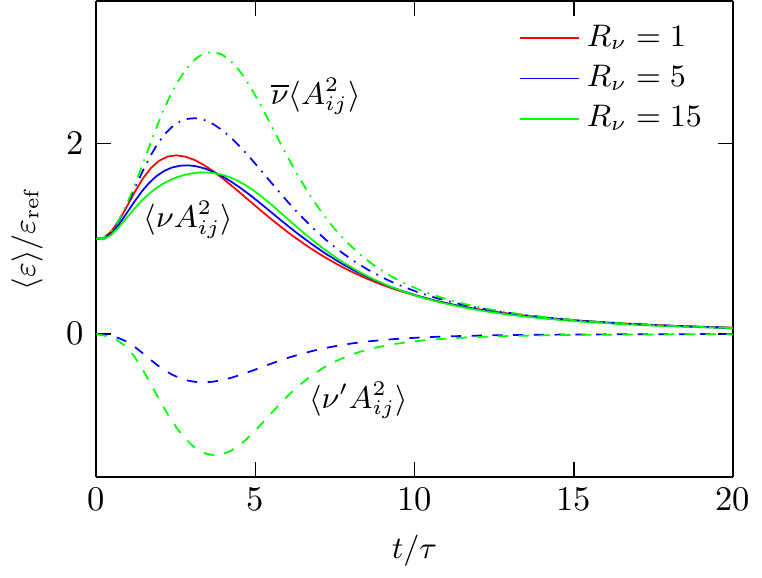} 
  \caption{Temporal evolution of the terms in eq.~\eqref{eq:ed}, i.e.\
    the mean dissipation rate
    $\avg{\varepsilon}_{\rm VV}=\avg{\nu A_{ij}^2}$ (solid lines), the
    dissipation formed with the mean viscosity
    $\avg{\varepsilon}_{\rm CV}=\avg \nu \avg{A_{ij}^2}$
    (dashed-dotted lines) and the fluctuating part
    $\avg{\varepsilon}_{\nu'}= \avg{\nu' A_{ij}^2}$ (dashed
    lines). The latter term is not shown for $R_\nu=1$, as it is zero
    by definition. The time is normalized by
    $\tau=1/(\kappa_p u'_{0})$. The reference dissipation is
    chosen as $\varepsilon_{\rm ref}=\avg{\varepsilon}(t=0)$.  }
  \label{fig:ed}
\end{figure}

The mean dissipation $\avg{\varepsilon}_{\rm VV}$ is defined as the
correlation between the local viscosity~$\nu$ and the square of the
local velocity gradient tensor
\begin{equation}
  A_{i j} = \pp{u_i}{x_j} \,.
\end{equation}
$\avg{\varepsilon}_{\rm VV}$
can be further decomposed by virtue of eq.~\eqref{eq:nu} as
\begin{equation}
  \label{eq:ed}
  \avg{\varepsilon}_{\rm VV} = \underbrace{\avg{\nu} \left\langle{\left( \pp{u_i}{u_j}
        \right)^2}\right\rangle}_{\varepsilon_{CV}} + \underbrace{\left\langle{\nu' \left( \pp{u_i}{u_j} \right)^2 }\right\rangle}_{\varepsilon_{\nu'}} \,.
\end{equation}
The first term $\varepsilon_{CV}$ on the right-hand side of
\eqref{eq:ed} denotes the classical dissipation for flows with
constant viscosity, while the second term $\varepsilon_{\nu'}$
accounts for dissipation due to fluctuations of the
viscosity. Figure~\ref{fig:ed} shows the temporal evolution of the
terms in \eqref{eq:ed} for the three cases under consideration. The
mean dissipation $\avg{\varepsilon}_{\rm VV}$ displays an initial
transient growth before turning into a decaying state, and is only
slightly affected by the viscosity ratio $R_\nu$. During the transient
period the increase of $\avg{\varepsilon}_{\rm VV}$ is delayed for the
two cases with variable viscosity.  During the early phase of the
decay, $\avg{\varepsilon}_{\rm VV}$ of the variable viscosity cases
exceeds the constant viscosity case, indicating an enhanced turbulent
mixing process.  At later times, when turbulent mixing advances and
the amplitude of viscosity fluctuations decreases, no difference
between the different cases can be discerned.  

Moreover, fig.~\ref{fig:ed} shows that different from
$\avg{\varepsilon}_{\rm VV}$, both $\avg{\varepsilon}_{\rm CV}$ and
$\avg{\varepsilon}_{\rm \nu'}$ strongly depend on the viscosity ratio
$R_\nu$. The dissipation $\avg{\varepsilon}_{\rm CV}$, which is built
with the mean viscosity $\avg\nu$, is positive and increases with
increasing viscosity ratio $R_\nu$ and clearly exceeds the mean
dissipation $\avg{\varepsilon}_{\rm VV}$. As the mean viscosity
$\avg\nu$ is the same for all cases, this indicates that turbulent
flows with variable viscosity are characterized by larger velocity
gradients $\avg{A_{ij}^2}$. In eq.~\eqref{eq:ed}, the strong
dependence of $\avg{\varepsilon}_{\rm CV}$ on $R_\nu$ is compensated
by the term $\avg{\varepsilon}_{\nu'}$ formed with the fluctuating
part of the viscosity $\nu'$. Initially, $\avg{\varepsilon}_{\nu'}$ is
zero because viscosity fluctuations and velocity gradients are
uncorrelated.  During the transient period, $\avg{\varepsilon}_{\nu'}$
becomes negative and tends to zero again for larger times.  Due to the
balance between the positive $\avg{\varepsilon}_{\rm CV}$ and the
negative $\avg{\varepsilon}_{\nu'}$, the mean dissipation
$\avg{\varepsilon}_{\rm VV}$ is only slightly depending on $R_\nu$.

\begin{figure}
  \centering
  \includegraphics[width=0.65\linewidth]{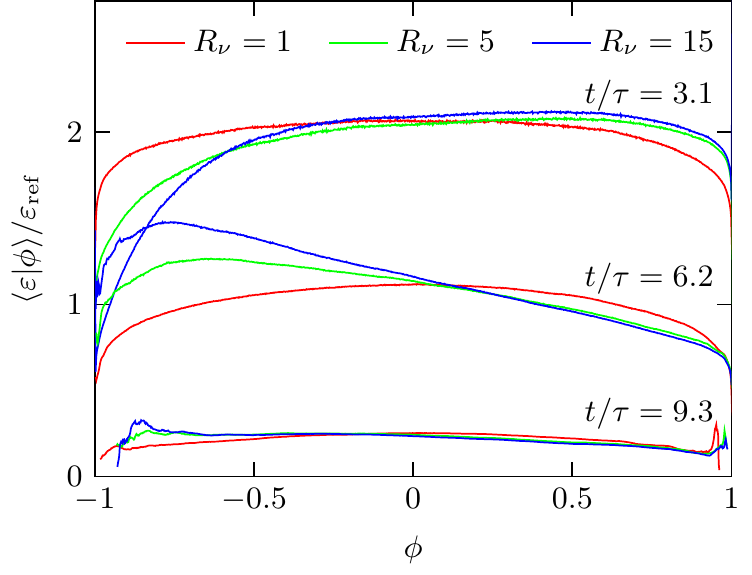} 
  \caption{Mean dissipation $\avg{\varepsilon|\phi}$ conditioned on
    the scalar $\phi$ for three different times. The reference
    dissipation is defined as
    $\varepsilon_{\rm ref}=u'^3_0\kappa_p$. }
  \label{fig:cmean}
\end{figure}

The conditional average $\avg{\varepsilon|\phi}$, shown in
fig.~\ref{fig:cmean}, provides further information on the correlation
between dissipation and viscosity. We discuss the temporal evolution
of the normalized conditional dissipation $\avg{\varepsilon|\phi}$ to
show how the dissipation adapts to viscosity as predicted by Taylor's
postulate, i.e.
\begin{equation}
  \label{eq:ed_phi}
  \avg{\varepsilon|\phi} \propto \frac{u'^3}{l_t} \,.
\end{equation}
The general expectation is that the dissipation is independent of
viscosity due to the fact that $l_t$ and $u'$ are large scale
quantities, which are virtually independent of fluctuations of the
viscosity. However, \cite{djenidi2017normalized} showed within the
framework of a scale-by-scale budget analysis that at finite Reynolds
numbers, eq.~\eqref{eq:ed_phi} is strictly valid only in
self-preserving turbulence.  Figure~\ref{fig:cmean} displays the
conditional dissipation $\avg{\varepsilon|\phi}$ for three different
time steps during the transition. A reference time is defined as
$\tau=1/(\kappa_p u'_{0})$. Shortly after initialization
($t/\tau=3.1$), the conditional dissipation $\avg{\varepsilon|\phi}$
is asymmetric for $R_\nu \ne 1$, signifying that the dissipation is
not independent of viscosity. At this time, large dissipation values
occur on average for $\phi>0$. At time step $t/\tau=6.2$, the
dissipation decays but the decay rate is more rapid in the
high-viscosity region (for $\phi>0$) and, hence, the asymmetry is
inverted and the highest dissipation level is observed in the low
viscosity regime for $\phi<0$. At later times ($t/\tau=9.3$), the
conditional dissipation $\avg{\varepsilon|\phi}$ decays further but
all curves virtually collapse and become independent of $\phi$. These
findings justify the validity of Taylor's postulate for turbulent
flows with variable viscosity. The normalized dissipation becomes
independent of viscosity after turbulence has adapted itself to the
fluctuating viscosity field. A similar observation was made by
\cite{lee2008validity}, by studying the turbulent mixing between two
initially segregated fluids with vastly different viscosities. They
found that during the mixing process, velocity gradients adapt rapidly
to the viscosity field, and in less than one-half integral time,
dissipation-viscosity independence was established.

\begin{figure}
  \centering
  \begin{tikzpicture}
    \draw (0, 0) node[inner sep=0]
    {\includegraphics[width=0.65\linewidth]{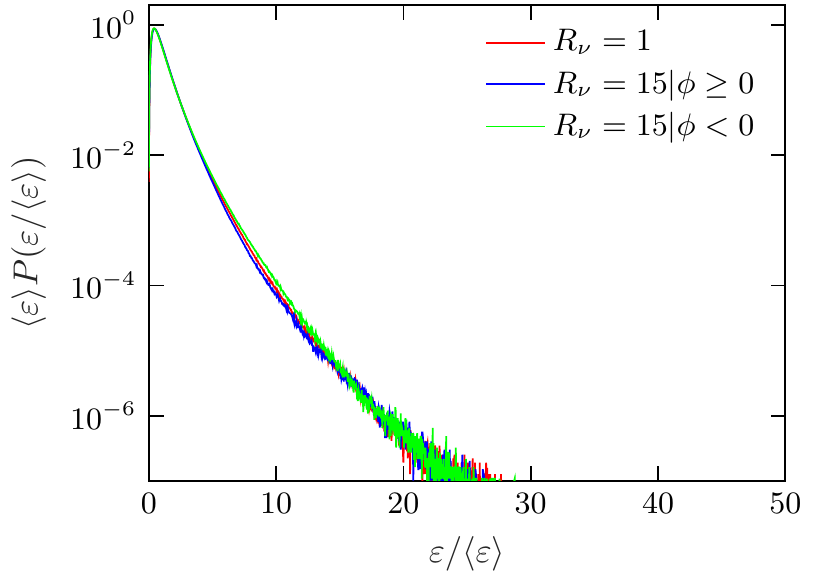}};
    \draw (0, 6.5) node[inner sep=0]
    {\includegraphics[width=0.65\linewidth]{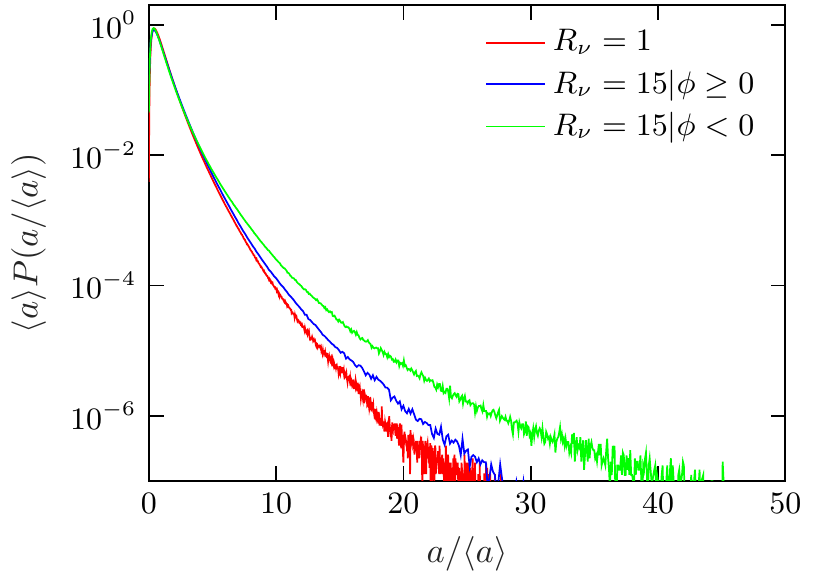}};
  \end{tikzpicture}
  \caption{Normalized marginal pdf of $a=A_{ij}^2$ (top) and
    $\varepsilon=\nu A_{ij}^2$ (bottom), conditioned on positive and
    negative values of $\phi$. Note that $\nu$ is linearly related to
    $\phi$ by eq.~\eqref{eq:nu}. The curves are normalized by the
    individual mean values $\avg{\varepsilon}$ and
    $\avg{a}=\avg{A_{ij}^2}$, respectively. The figure shows the case
    $R_\nu=15$ at $t/\tau=9.3$.  For reference, the constant
    viscosity case $R_\nu=1$ is shown.}
  \label{fig:mpdf}
\end{figure}

Let us now analyze the link between $\nu$ and $A_{ij}^2$ in more
detail by probability density functions (pdf). To do this, we focus on
the time step $t/\tau=9.3$, at which the dissipation has already
adapted itself to the fluctuating viscosity field. This makes our
analysis independent from transitional effects and reduces the impact
of the initial conditions. An important feature of $A_{ij}^2$ (and the
dissipation field) is the presence of large fluctuations, which exceed
the respective mean values by orders of magnitude.  This phenomenon is
known in literature as internal intermittency,
cf.~\cite{frisch1995turbulence}.   Generally, the term
  intermittency describes a random process, which exhibits very strong
  events that occur more frequently than predicted by a Gaussian
  distribution. Following \cite{nelkin1994universality}, the origin of
  intermittency lies in the non-linear dynamics of the vortex
  stretching mechanism. For variable viscosity turbulence, another
  source of intermittency can be identified by the non-linearity of
  the viscous term. To address this assertion, fig.~\ref{fig:mpdf}
  illustrates the normalized pdfs of $A_{ij}^2$ and
  $\varepsilon=\nu A_{ij}^2$ conditioned on $\phi\ge 0$ and $\phi<0$,
  respectively (note that $\nu$ is linearly related to $\phi$ by
  eq.~\eqref{eq:nu}). The normalized pdf of $A_{ij}^2$ has stretched
  exponential tails, which originate from strong rare events that are
  non-universal and generally depend on Reynolds number. The far tails
  contribute mostly to higher-order moments and can be used to measure
  intermittency. From fig.~\ref{fig:mpdf}, it can be observed that the
  tail of the $A_{ij}^2$ pdf in the low viscosity regime is
  significantly pronounced compared to the high viscosity regime. For
  comparison, the pdf of the constant viscosity case is shown. It has
  a less pronounced tail, compared to both low and high viscosity
  regimes of the variable viscosity case. This observation suggests
  that variable viscosity turbulence exhibits stronger fluctuations of
  the velocity gradients compared to constant viscosity turbulence at
  the same mean viscosity $\avg{\nu}$. This allows the conclusion that
  variable viscosity turbulence is characterized by an enhanced level
  of intermittency. A similar conclusion was drawn by
  \citet{grea2014effects} based on an EDQNM closure model.  With this
  closure, they predicted that a variable viscosity fluid behaves like
  a fluid with constant but lower average viscosity, and therefore
  higher intermittency.

\begin{figure}
  \centering
  \includegraphics[width=0.65\linewidth]{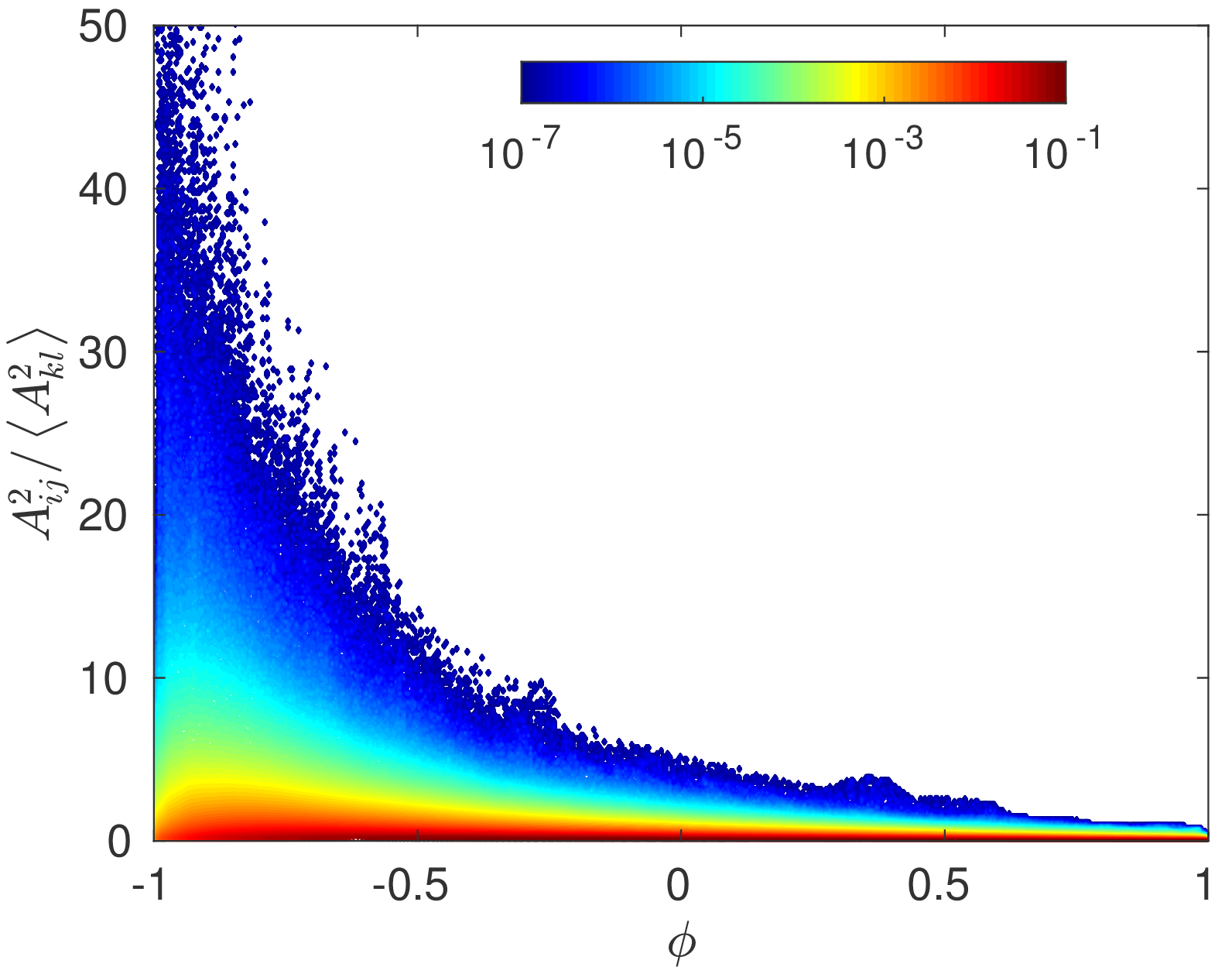}
  \caption{Joint pdf $P(A_{ij}^2,\phi)$ indicates the presence of
    large velocity gradients especially in the low viscosity regime
    for $\phi<0$. The figure shows the case $R_\nu=15$ at
    $t/\tau=9.3$. The color is scaled logarithmically and increases in
    the sequence blue-green-yellow-red.}
  \label{fig:jpdf}
\end{figure}

  The normalized conditional pdf of the dissipation reveals a notably
  reduced dependence on viscosity, which signifies that statistics of
  the dissipation $\avg{{\varepsilon}}_{\rm VV}$ become independent of
  viscosity. This finding can be explained by the correlation $\nu$
  and $A_{ij}^2$.  Figure~\ref{fig:jpdf} shows that large velocity
  gradients occur on average at low viscosity (and vice versa). This
  negative correlation between $A_{ij}^2$ and $\nu'$, i.e.\
\begin{equation}
\avg{\nu' A_{ij}^2}<0\,,
\end{equation}
is a necessary condition for the independence of the dissipation from
viscosity.

\section{Impact of viscosity fluctuations on the viscous cut-off
  length scale}

\begin{figure}
  \centering
  \includegraphics[width=0.65\linewidth]{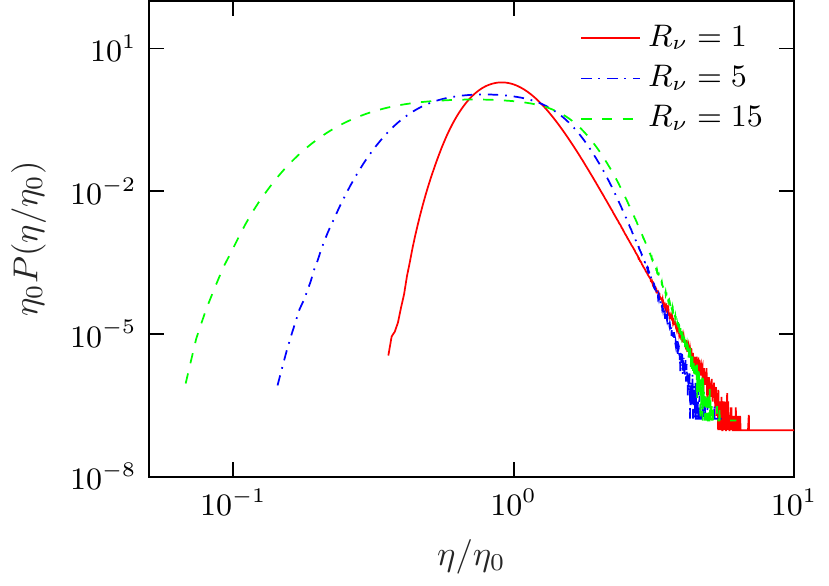}
  \caption{Normalized probability density function of the Kolmogorov
    length scale $\eta=(\nu^3/\varepsilon)^{1/4}$ at $t/\tau=9.3$.}
  \label{fig:eta}
\end{figure}

Kolmogorov's scaling theory for turbulence
(\cite{kolmogorov1941dissipation,kolmogorov1941local}) postulates the
existence of a dissipative cut-off scale where the turbulent cascade
ends.  This length scale is known as the Kolmogorov length
  scale
\begin{equation}
  \eta_0 = \left( \frac{\avg{\nu}^3}{\avg{\varepsilon}} \right)^{1/4} \,.
\end{equation}
In a turbulent flow, high values of the local dissipation
$\varepsilon$ occur around thin sheet or tube-like regions, which are
characterized by length scales much smaller than the Kolmogorov length
scale $\eta_0$ built with the mean viscosity $\avg{\nu}$ and mean
dissipation $\avg{\varepsilon}$. To study fluctuations of the local
dissipative cut-off scale, we follow \cite{schumacher2007sub} and
introduce a local Kolmogorov length, defined as
\begin{equation}
  \eta = \left(  \frac{{\nu}^3}{{\varepsilon}}\right)^{1/4}, 
\end{equation}
which is itself a fluctuating quantity.  To quantify variations of
 the local Kolmogorov length $\eta$, the probability density
  function $P(\eta)$ is of interest.  The normalized pdf of $\eta$ is
  displayed in fig.~\ref{fig:eta} for all cases, and shows that
  generally, length scales much smaller than $\eta_0$ exist. The left
tail of the pdf, which is dominated by large intermittent fluctuations
of the dissipation, reveals a strong dependence on the viscosity ratio
$R_\nu$. The cases with variable viscosity show a clear tendency to
establish significantly smaller cut-off scales than the case with
constant viscosity.  On the other hand, the right tail of the pdf is
nearly unaffected by viscosity fluctuations. These findings indicate
that particularly the strong intermittent events of turbulence are
enhanced by the fluctuations of viscosity, resulting in the presence
of smaller length scales and a modified turbulent mixing. A further
justification of these results is provided by the joint pdf of the
velocity gradients and the viscosity $P(A_{ij}^2,\phi)$, where the
largest velocity gradients $A_{ij}^2$ occur in the regime of low
viscosity, see fig.~\ref{fig:jpdf}. This observation signifies that
the smallest length scales occur as expected in regions of low
viscosity.

\begin{figure}
  \centering
  \includegraphics[width=0.75\linewidth]{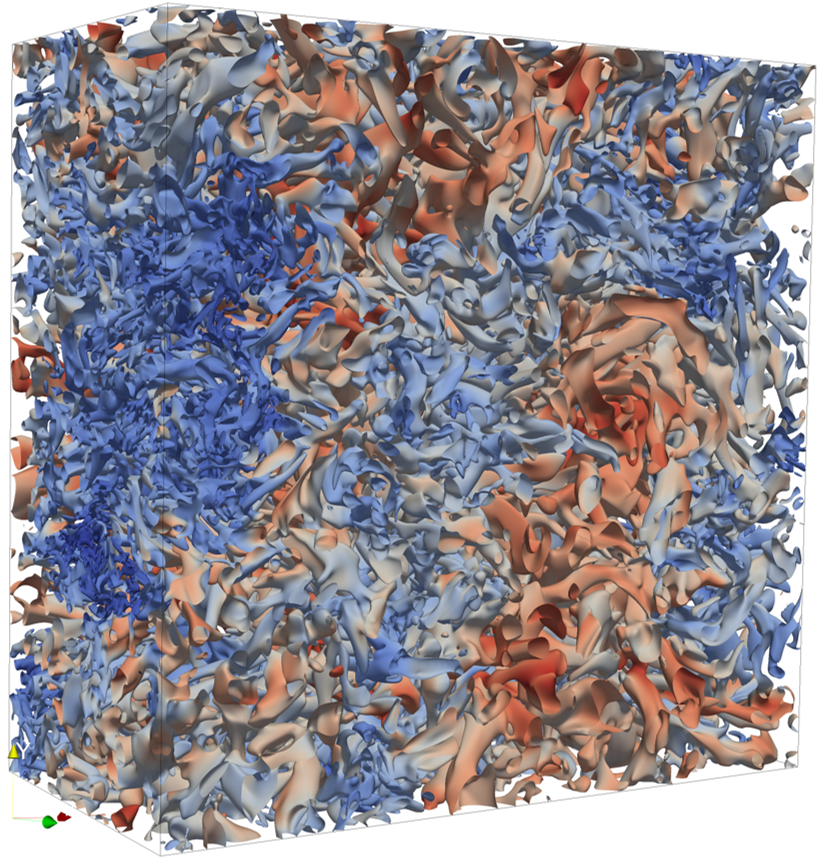}
  \caption{Iso-surface of vorticity defined by the Q-criterion,
    cf.~\cite{dubief2000coherent}, for case $R_\nu=15$. The
    iso-surface is colored by local viscosity, where the color scale
    increases in the sequence blue-white-red.}
  \label{fig:q}
\end{figure}

To illustrate the impact of viscosity fluctuations on the local
structure of turbulence, we display in fig.~\ref{fig:q} iso-surfaces
of the vorticity, defined by the Q-criterion
\begin{equation}
  \label{eq:q}
  Q = \frac{1}{2} \left( \omega_{ij}^2 - s_{ij}^2 \right),
\end{equation}
where $\omega_{ij}$ and $s_{ij}$ are the anti-symmetric and symmetric
components of the velocity gradient tensor, respectively,
cf.~\cite{dubief2000coherent}. The iso-surfaces are colored by local
viscosity. The turbulent field, defined by \eqref{eq:q}, is
characterized by a high level of intermittency. Figure~\ref{fig:q}
reveals an enormous number of degrees of freedom represented by the
interaction of vortices of different size that tend to cluster in
coherent structures.  A correlation between the local viscosity and
the size of the vortex structures is clearly visible. Regions of low
viscosity exhibit a fine vortex structure, while the size of vortex
structures is considerably larger in regions of high viscosity. The
reason for this observation is the damping effect of viscosity
resulting in different local Reynolds numbers.

\section{ Scale-by-scale budget equation for turbulence with variable
  viscosity}

A central quantity of turbulence research is the velocity increment
$\Delta u=u(\vec x + \vec r) - u(\vec x)$, defined as the velocity
difference between two points separated by a distance $\vec r$.  The
moments of the velocity increment are known as structure functions.
Transport equations for structure functions have been widely studied
in literature and can be derived from first principles,
cf.~\cite{danaila2001turbulent} and \cite{gauding2014generalised}.
Structure functions can be used to analyze scale-dependent properties,
since they capture not only local but also non-local phenomena. 
  We derived a novel transport equation for the so-called energy
  structure function $\avg{(\Delta u_j)^2}$ for variable viscosity
  fluids.  For statistically homogeneous incompressible turbulence,
  this equation reads
\begin{equation}
  \label{eq:sbs}
\begin{aligned}
\pp{}{t}\avg{(\Delta u_j)^2} + &\pp{}{r_i} \left\langle \Delta u_i (\Delta
  u_j)^2 \right\rangle = 
\pp{^2}{r_i^2} \avg{(\nu + \nu^+)(\Delta u_j)^2} \;    \\
-&2 \left\langle \nu \left(\pp{u_j}{x_i} \right)^2 + {\nu^+
    \left(\pp{u_j^+}{x_i^+} \right)^2} \right\rangle  \\
-&\pp{}{r_i} \left\langle \left( \pp{\nu^+}{x_i^+} - \pp{\nu}{x_i} \right)
  (\Delta u_j)^2 \right\rangle\; \\ +& 
\left\langle \Delta u_j \left( \pp{\nu^+}{x_i^+}\pp{u_i^+}{x_j^+} -
    \pp{\nu}{x_i}\pp{u_i}{x_j}  \right)\right\rangle\; .
\end{aligned}
\end{equation}
The general procedure to derive eq.~\eqref{eq:sbs} is to formulate the
Navier-Stokes equations at two different and independent points
$\vec x$ and~$\vec x^+$. For notational clarity, the super-script $+$
refers to quantities defined at the point $\vec x^+=\vec x+ \vec r$.
Subtracting the governing equations at these two points leads to an
evolution equation for the velocity increment $\Delta u_j$. 
  Multiplication of the result with $2\Delta u_j$ and
  ensemble-averaging gives eq.~\eqref{eq:sbs}. Due to statistical
  homogeneity of the considered flow, the ensemble-averaged equation
  is independent of position~$\vec x$, and depends solely on the
  separation vector~$\vec r$.

Equation \eqref{eq:sbs} is a generalized scale-by-scale budget
equation for the turbulent energy at scale~$r$ for turbulent flows
with variable viscosity. The terms on the left-hand side represent
unsteady effects and turbulent inter-scale transport. The first two
terms on the right-hand side represent molecular inter-scale transport
and dissipation of turbulent energy. The last two terms on the
right-hand side occur only in turbulence with variable viscosity. They
result from the viscous term of eq.~\eqref{eq:ns} and represent
inter-scale transport and molecular transport due to viscosity
gradients.  A scale-by-scale budget equation for variable
  viscosity turbulence was derived before by
  \cite{voivenel2016similarity} without discriminating between
  transport induced by turbulence and transport induced by viscosity
  gradients.  

\section{ Analysis of scale-by-scale budget equations with constant and
  variable viscosity}

In the following, we will analyze the DNS based on structure
functions. First, we examine the longitudinal second-order velocity
structure function $\avg{(\Delta u)^2}$ as it appears in the unsteady
term. Then, the longitudinal mixed viscosity-velocity structure
function $\avg{(\nu +\nu^+)(\Delta u)^2}$ as it occurs in the
diffusive term of eq.~\eqref{eq:sbs} will be discussed. After that, we
turn our attention to the full budget as given by eq.~\eqref{eq:sbs}
to analyze the impact of viscosity gradients on the inter-scale
transport mechanism.

\begin{figure}
  \centering
  \includegraphics[width=0.65\linewidth]{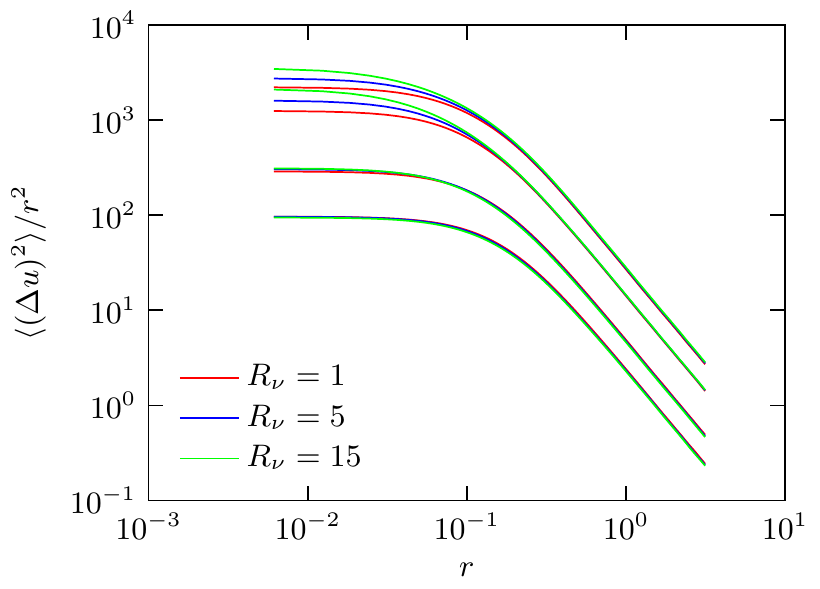} \\
  \includegraphics[width=0.65\linewidth]{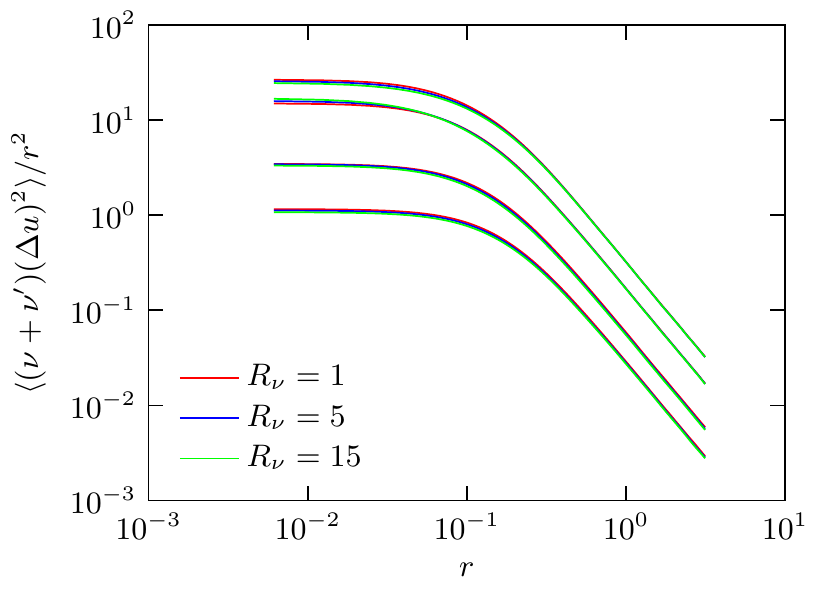}
  \caption{Compensated second order velocity structure function (top)
    and compensated mixed viscosity-velocity structure function
    (bottom) for different viscosity ratios $R_\nu$ and different
    times $t/\tau$ (the time steps $t/\tau=2.0$ (top-most), 3.1, 6.2,
    9.3 (bottom-most) are displayed).}
  \label{fig:sf}
\end{figure}

In the limit $r\to 0$, the second order structure function
$\avg{(\Delta u)^2}$ can be developed in a Taylor series, i.e.
\begin{equation}
  \label{eq:du2}
  \avg{(\Delta u)^2} = \left\langle \left( \pp{u}{x} \right)^2 \right\rangle r^2 
  = \frac{1}{15} \frac{\avg{\varepsilon}_{\rm CV}}{\avg\nu} r^2 \,,
\end{equation}
and is uniquely determined by $\avg \nu$ and
$\avg{\varepsilon}_{\rm CV}$.  In the derivation of eq.~\eqref{eq:du2}
it is assumed without loss of generality that the separation vector
$\vec r$ is aligned with the $x$ axis.  In the second equality the
factor $1/15$ is obtained by relating $\avg{\varepsilon}_{\rm CV}$ to
$\avg{(\partial u/\partial x)^2}$ due to isotropy. In the large-scale
limit for $r \to \infty$, the second order structure function becomes
independent of $r$ and tends to its one-point limit, which equals
$2\avg{u^2}$. Figure~\ref{fig:sf} shows the velocity structure
function $\avg{(\Delta u)^2}$ for different times, for the constant
and variable viscosity cases under consideration. The structure
functions are compensated by the dissipative range scaling $r^2$,
which yields a plateau for $r\to 0$. A dependence of the structure
function on the viscosity ratio is observed in the dissipative range
for the time steps $t/\tau=2.0$ and $t/\tau=3.1$, which is in
agreement with the dependence of $\avg{A_{ij}^2}$ on $R_\nu$. At later
times or at larger scales, $\avg{(\Delta u)^2}$ is independent of
$R_\nu$, which is in agreement with the independence of $\avg{k}$
on~$R_\nu$.

In a next
step, we consider the viscosity-velocity structure function, defined
as
\begin{equation}
  \label{eq:sf_nu}
  S_{\nu}=\avg{  (\nu(\vec x + \vec r) + \nu(\vec x)) (u(\vec r + \vec r) -
    u(\vec x))^2 } \,,
\end{equation}
which appears in the diffusive term of the transport equation for
$\avg{(\Delta u)^2}$ in variable viscosity turbulence,
cf.~eq.~\eqref{eq:sbs}. Equation \eqref{eq:sf_nu} can be developed in
a Taylor-series for $r\to 0$, yielding
\begin{equation}
  S_\nu= 2
  \left\langle\nu \left( \pp{u}{x} \right)^2 \right\rangle r^2  \,.
\end{equation}
In the large-scale limit $r \to \infty$, $S_\nu$ becomes independent
of $r$ and tends to $4\avg{\nu u^2}$. Under the assumption that the
viscosity is statistically independent from the turbulent energy,
$4\avg{\nu u^2}$ simplifies to $4\avg{\nu}\avg{u^2}$.  The
viscosity-velocity structure functions are shown in
fig.~\ref{fig:sf}. Different to $\avg{(\Delta u)^2}$, the
viscosity-velocity structure functions collapse for all time steps
under consideration over all scales $r$ independently of $R_\nu$. This
result underlines that large-scale statistics in statistical
homogeneous isotropic turbulence are not affected by fluctuations of
the viscosity due to the statistical independence between $u^2$ and
$\nu$.

\begin{figure}
  \centering
  \includegraphics[width=0.65\linewidth]{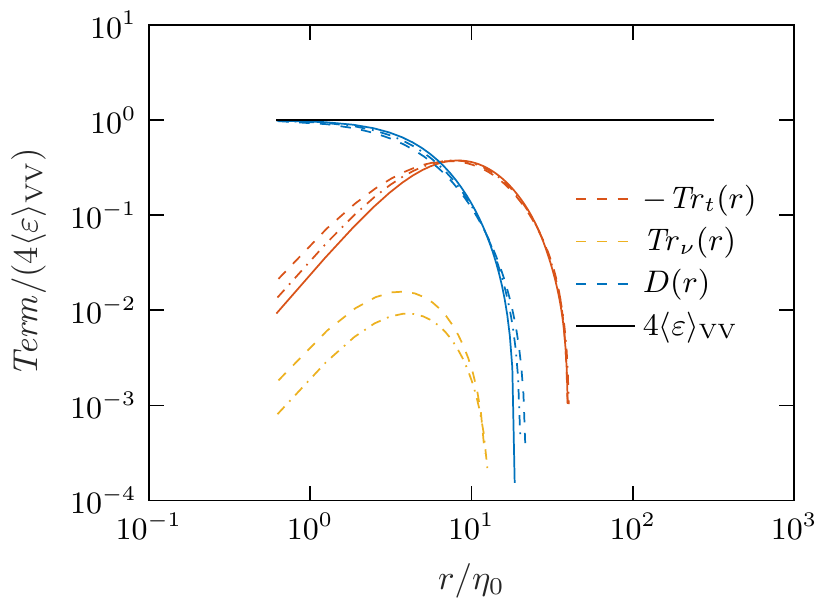} 
  \caption{Scale-by-scale budget according to eq.~\eqref{eq:sbs}
    showing the viscous term $D(r)$, the turbulent inter-scale
    transport term $T_t(r)$, and the inter-scale transport term due to
    viscosity gradients $T_\nu(r)$ for different viscosity ratios
    $R_\nu=1$ (solid lines), $R_\nu=5$ (dashed lines), and $R_\nu=15$
    (dashed-dotted lines). The scale-by-scale budget can be written
    with eqs.~\eqref{eq:D} - \eqref{eq:Trnu} as
    $\pp{}{t} \avg{(\Delta u_j)^2} - \mathit{Tr}_\nu - \mathit{Tr}_t +
    D = 4 \avg{\varepsilon}_{\rm VV}$. }
  \label{fig:sf3}
\end{figure}

Provided that the Reynolds number is very large and with the
assumptions of local isotropy and uniform viscosity,
eq.~\eqref{eq:sbs} simplifies in inertial range for
$\eta \gg r \gg l_t$ to
\begin{equation}
  - \avg{\Delta u_L (\Delta u_j)^2} = \frac{4}{3} \avg{\varepsilon} r \,,
\end{equation}
where $\Delta u_L$ being the longitudinal velocity increment aligned
with the separation vector $\vec r$, cf.~\cite{antonia1997analogy}. At
low Reynolds numbers, a well established inertial range does not exist
and we can generally not assume that inertial range statistics are
independent of viscosity fluctuations. In this case, we need to
consider all terms of eq.~\eqref{eq:sbs}. The scale-by-scale budget,
normalized by $4\avg{\varepsilon}_{\rm VV}$, is shown in
fig.~\ref{fig:sf3} for different $R_\nu$.  It can be observed, that in
the dissipative range the viscous terms $D(r)$, defined by
\begin{equation}
    \label{eq:D}
    D(r) = \pp{^2}{r_i^2} \avg{(\nu + \nu^+)(\Delta u_j)^2} +  \avg{\Delta u_j \left( \pp{\nu^+}{x_i^+}\pp{u_i^+}{x_j^+} -
        \pp{\nu}{x_i}\pp{u_i}{x_j}  \right)}  \,,
\end{equation}
are relevant and balance in the limit $r\to 0$ the dissipation, i.e
\begin{equation}
\lim_{r\to 0} D(r) = 4 \avg{\varepsilon}_{\rm VV}  \;.
\end{equation}

At intermediate scales, $\eta \ll r \ll l_t$, two different
inter-scale transport mechanisms exist, namely turbulent inter-scale
transport
\begin{equation}
  \mathit{Tr}_t = \pp{}{r_i} \avg{\Delta u_i (\Delta u_j)^2} \,,
\end{equation}
and inter-scale transport due to viscosity gradients 
\begin{equation}
    \label{eq:Trnu}
    \mathit{Tr}_\nu = \pp{}{r_i} \avg{\left( \pp{\nu^+}{x_i^+} - \pp{\nu}{x_i} \right)
      (\Delta u_j)^2} \,.
\end{equation}
As shown in fig.~\ref{fig:sf3}, the inter-scale transport due to
viscosity $\mathit{Tr}_\nu$ is small compared to the turbulent
inter-scale transport $\mathit{Tr}_t$. Due to the relatively low
Reynolds number, the flow does not exhibit a clear inertial range. The
turbulent transport~$\mathit{Tr}_t$ is negative, which confirms the
well known forward cascade of turbulence with a mean transport of
turbulent energy from the large scales towards the small
scales. However,~$\mathit{Tr}_\nu$ is positive implying that viscosity
gradients induce an inverse  transport, where energy propagates
  from the small scales to the large scales. This finding is
important because it signifies that variable viscosity effects are not
only limited to small scales, but can also affect larger scales as
found by \cite{voivenel2017variable} and \cite{danaila2017self} by
studying the morphology of turbulent jet flows with variable
viscosity. Similar observations have been made in the context of
differential diffusion.  Differential diffusion has a molecular
  origin and also reveals an inverse transport of energy from the
  small scales to the large scales, cf.~\cite{yeung1996multi} and
  \cite{hunger2016impact}.

\section{Summary and discussion}
In the present study, we analyzed homogeneous isotropic decaying
turbulence with variable viscosity at low Reynolds number by means of
highly resolved direct numerical simulations. Three different
viscosity ratios between 1 and 15 were considered.

The main results are: \\
(i) An analysis of the transport equation for the turbulent kinetic
energy showed that the effect of variable viscosity is virtually
negligible on both the mean turbulent energy $\avg{k}$ and the mean
dissipation $\avg{\varepsilon}_{\rm VV}$.  This finding is in
  agreement with previous observations by \cite{lee2008validity} and
  \cite{grea2014effects}. It confirms the validity of Taylor's
  postulate, which states that the (normalized)
  dissipation is independent of viscosity. \\
(ii) Turbulent flows with variable viscosity reveal significantly
enhanced velocity gradients in regions of low viscosity, which results
in the presence of smaller length scales and an increased level of
small-scale intermittency. \\
(iii) A generalized scale-by-scale budget equation for the energy
structure function $\avg{(\Delta u_j)^2}$ in variable viscosity flows
was derived. A new term appears in this equation that accounts for an
additional inter-scale transport due to viscosity gradients.  This
contribution is small compared to the turbulent inter-scale
transport.  But different to the turbulent inter-scale
  transport, the viscosity gradient induced transport is directed from
  the small scales to the large scales. 

%


\section*{Acknowledgment}
Financial support was provided by the Labex EMC3, under the grant
VAVIDEN, as well as the Normandy Region and FEDER. Additionally, the
authors gratefully acknowledge the computing time granted on the
supercomputer JUQUEEN (Research Center Juelich,
cf.~\cite{stephan2015juqueen}), the supercomputer TURING (IDRIS), and
the supercomputer Myria (CRIANN).


\section*{References}
\bibliographystyle{elsarticle-harv} 
\bibliography{tsfp}

\end{document}